# ERK/p38 MAPK inhibition reduces radio-resistance to pulsed proton beam in breast cancer stem cells

Myung-Hwan Jung,* Jeong Chan Park

Korea Multi-purpose Accelerator Complex, Korea Atomic Energy Research Institute, Gyeong-ju 780-904

Recent studies have identified highly tumorigenic cells with stem cell-like characteristics in human cancers, termed cancer stem cells (CSCs). CSCs are resistant to conventional radiotherapy and chemotherapy owing to their high DNA repair ability and oncogene overexpression. However, the mechanisms regulating CSC radio-resistance, particularly proton beam resistance, remain unclear. We isolated CSCs from the breast cancer cell lines MCF-7 and MDA-MB-231, which expressed the characteristic breast CSC membrane protein markers $CD44^{+}/CD24^{-/low}$, and irradiated the CSCs with pulsed proton beams. We confirmed that CSCs are resistant to pulsed proton beams and showed that treatment with p38 and ERK inhibitors reduced CSC radio-resistance. Based on these results, BCSC radio-resistance can be reduced during proton beam therapy by co-treatment with ERK1/2 or p38 inhibitors, representing a novel approach for breast cancer therapy.

Email: jungmh80@kaeri.re.kr

Fax: +82-54-705-3738

# I. INTRODUCTION

Breast cancer is the most common cancer in women and the second most common cancer in the world [1]. Breast cancer is primarily treated by surgery, with response rates ranging from 60% to 80% for primary breast cancer. However, local control fails in 20–70% of patients within 5 years [2]. Cancer recurrence is attributed to cancer stem cells (CSCs) in malignant tumors.

CSCs possess characteristics associated with normal stem cells: They can self-replicate (self-renewal) and differentiate into multiple cell types [3]. Furthermore, they are resistant to conventional radiotherapy, such as photon beam irradiation, and chemotherapy owing to high DNA repair ability and oncogene overexpression [4]. However, the mechanisms regulating CSC radio-resistance, including to proton beam, remain unclear.

Breast cancer stem cells (BCSCs) were first detected by Al-Hajj *et al.* [5]. They showed that a subset of cells expressing CD44 with weak or no CD24 expression could establish new tumors in xenograft mice. Recently, BCSC-targeting therapies have been evaluated by numerous groups. Strategies include targeting BCSC self-renewal [6], indirectly targeting the microenvironment [7,8], and directly killing BCSCs by chemical agents that induce differentiation [9-12], immunotherapy [13], and oncolytic viruses [14].

Mitogen-activated protein kinases (MAPKs) are protein kinases that direct cellular responses to diverse stimuli, including mitogens, osmotic stress, heat shock, and proinflammatory cytokines. They also regulate cell proliferation, differentiation, cell survival, apoptosis, gene expression, and motility [15]. Extracellular signal-regulated kinases 1 and 2 (ERK1/2), c-Jun amino-terminal kinase (JNK), and p38 are well-known MAPKs. ERK1/2 and JNK play a crucial role in cell proliferation via the induction of positive cell cycle regulators [16] and AP-1 complex formation and AP-1-mediated gene transcription [17], respectively. In contrast, p38 plays an important role in inflammatory responses [18].

The present study evaluated whether CSCs are resistant to pulsed proton beam irradiation and the effect of MAPK inhibitor treatment on CSC radio-resistance.

## II. MATERIALS AND METHODS

### 1. BCSC isolation

MCF-7 and MDA-MB-231 cells ($1 \times 10^7$ cells/ml) were maintained in phosphate buffed saline (PBS) containing 0.5% bovine serum albumin (BSA). Fluorochrome-conjugated monoclonal antibodies against FITC-CD44 and PE-conjugated CD24 antibodies (BD Bioscience, Franklin Lakes, NJ, USA), or their respective IgG controls, were added to the cell suspension according to the manufacturer's protocol and incubated at room temperature in the dark for 0.5 to 1 h. The labeled cells were washed with wash buffer, then sorted on a FACS Jazz (BD Biosciences) to obtain $CD44^+/CD24^{-/low}$ cells.

### 2. Pulsed proton beam irradiation

Cells were irradiated with pulsed proton beams at the Korea Multi-purpose Accelerator Complex (KOMAC). Protons (45 MeV) were produced with a spread Bragg peak using a ridge filter modulator [19]. Cells were irradiated with a single dose (4, 8, 10, or 12 Gy) of protons. The average dose rate was 1 Gy/pulse. To monitor the dose, we used radio-chromic film (MD-V3, Ashland, Dublin, OH, USA). Attached cells were irradiated in a 96-well plate or T-25 flask filled with medium and placed on a beam stage. Plates were oriented so that the growth surface was orthogonal to the horizontal beam entering the plate bottom.

### 3. Proliferation assay

MCF-7 and MDA-MB-231 cells ($1 \times 10^3$ cells/well) were plated in RPMI 1640 medium supplemented with 10% FBS for 24 h, followed by irradiation with pulsed proton beams (4 to 12 Gy). Cells were further cultured for 3 days. Cell viability was determined using the Cell Counting Kit-8 (CCK-8, Dojindo, Tokyo, Japan) and a microplate reader at 450 nm. Each condition was tested in quadruplicate, and the fold induction compared to the control cells was determined.

**4. Colony formation assay**

To evaluate radio-resistance, MCF-7 and MDA-MB-231 cells were incubated for 24 h and plated in T-25 culture flasks (400 cells/well). Cells were irradiated with pulsed proton beams at a single dose of 4 or 10 Gy. After 9 days, the media was removed from the wells. Colonies were washed twice with PBS and fixed with methanol (5 min). Cells were then washed with distilled water and stained with methylene blue solution. The surviving fraction was calculated as the number of colonies formed, and the fold induction compared to the control cells was determined.

**5. Inhibition of MAPKs**

To evaluate the effects of MAPK inhibition, we pre-treated cells with the MAPK inhibitors, purchased from Sigma Aldrich, PD98059 (ERK1/2 inhibitor, 2 μM), SB203580 (p38 inhibitor, 2 μM), and SP600125 (JNK inhibitor, 2 μM) for 3 h.

## III. RESULTS

**1. BCSCs are resistant to pulsed proton beam irradiation.**

To evaluate BCSC and normal breast cancer cell (BCC) resistance to pulse proton beam irradiation, we separated the cells using FACS. FACS analysis revealed that BCSCs comprise 17% of MDA-MB-231 cells (Fig. 1A). In contrast, less than 1% of MCF-7 cells are BCSCs (Fig. 1B). Pulsed proton beam irradiation experiments (4 to 12 Gy) were carried out in BCSCs and BCCs after cell sorting. First, cell growth was analyzed using a proliferation assay. Although pulsed proton beam irradiation reduced survival in MDA-MB-231 BCSCs and BCCs after 72 h, there was no statistically significant difference between the two cell types (Fig. 2A). In contrast, MCF-7 BCC growth was dose-dependently inhibited by pulsed proton beams. However, MCF-7 BCSCs were resistant to radio-therapy (Fig. 2B)

To further investigate BCSC and BCC resistance to pulsed proton beam-induced cell death, a colony formation assay was performed. The results of colony formation assay were similar to the results of CCK-8 assay. As shown in Figure 3A, there was no significant between the BCSCs and BCCs in MDA-MB-231 cells. In contrast, MCF-7 BCSCs were 2-3 fold more resistant to proton beam-induced cell death than MCF-7 BCCs (Fig. 3B)

**2. MAPK inhibition reduced radio-resistance in BCSCs.**

As shown in Figures 2 and 3, MCF-7 BCSCs had the highest radio-resistance to pulsed proton beam irradiation. This suggests that BCSC-mediated cancer relapse could occur following proton therapy. We investigated whether MAPK inhibition could alter BCSC resistance to pulsed proton beam irradiation. BCSCs and BCCs in MCF-7 and MDA-MB-231 cell lines were treated with PD98059 (ERK1/2 inhibitor), SB203580 (p38 inhibitor), and SP600125 (JNK inhibitor), and cell survival was determined following pulsed proton beam irradiation. Pulsed proton beam radio-resistance was reduced in BCSCs following treatment with PD98059 or SB203580 (Fig. 4).

## IV. CONCLUSION

In the current study, we showed that BCSCs are resistant to pulsed proton beam irradiation (Fig. 2 and 3) and that MAPK inhibitors specifically targeting ERK1/2 or p38 MAPK could overcome this radio-resistance (Fig. 4). Reducing BCSC resistance to pulsed proton beams is essential to improve therapeutic efficacy and decrease the 5-year recurrence rate. Based on these results, BCSC radio-resistance can be reduced during proton beam therapy by co-treatment with ERK1/2 or p38 inhibitors, representing a novel approach for breast cancer therapy.

**ACKNOWLEDGMENT**

This work has been supported through KOMAC (Korea of Multi-purpose Accelerator Complex) operation fund of KAERI by MSIP(Ministry of Science, ICT and Future Planning)

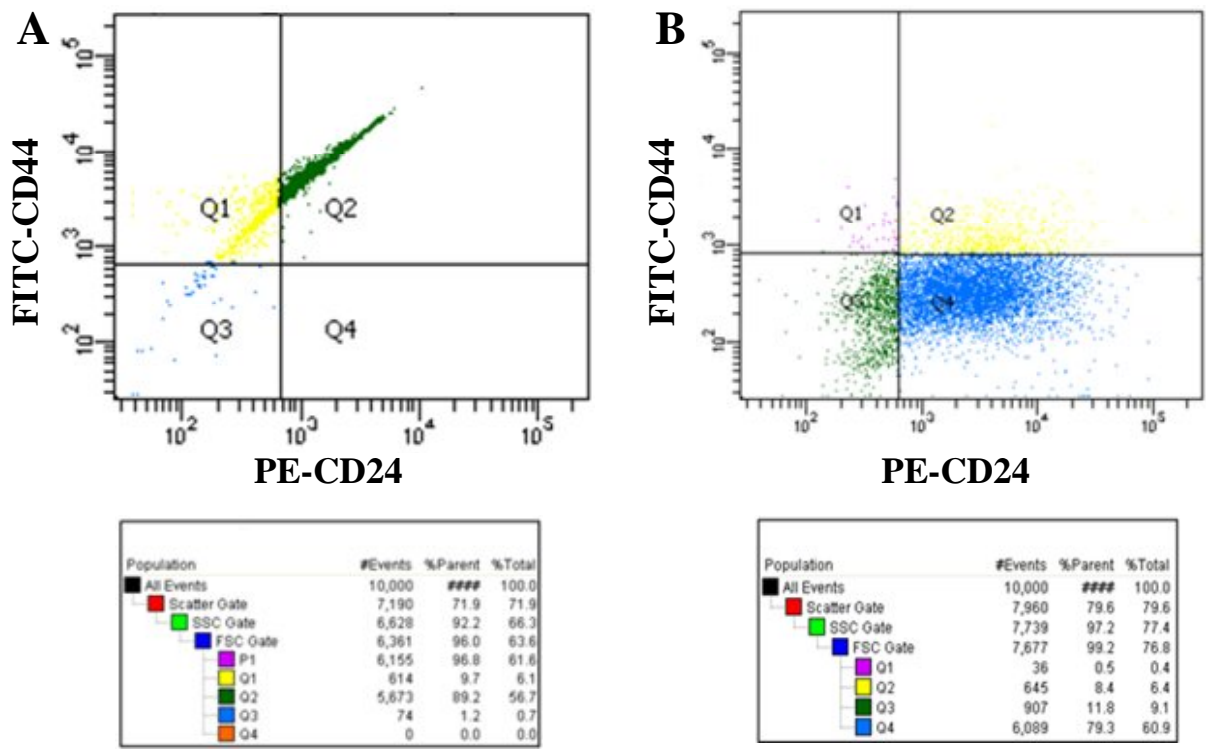

Fig. 1. The Results of BCSCs isolation by using fluorescence-activated cell sorting. (A) MDA-MB-231 cell line, (B) MCF-7 cell line

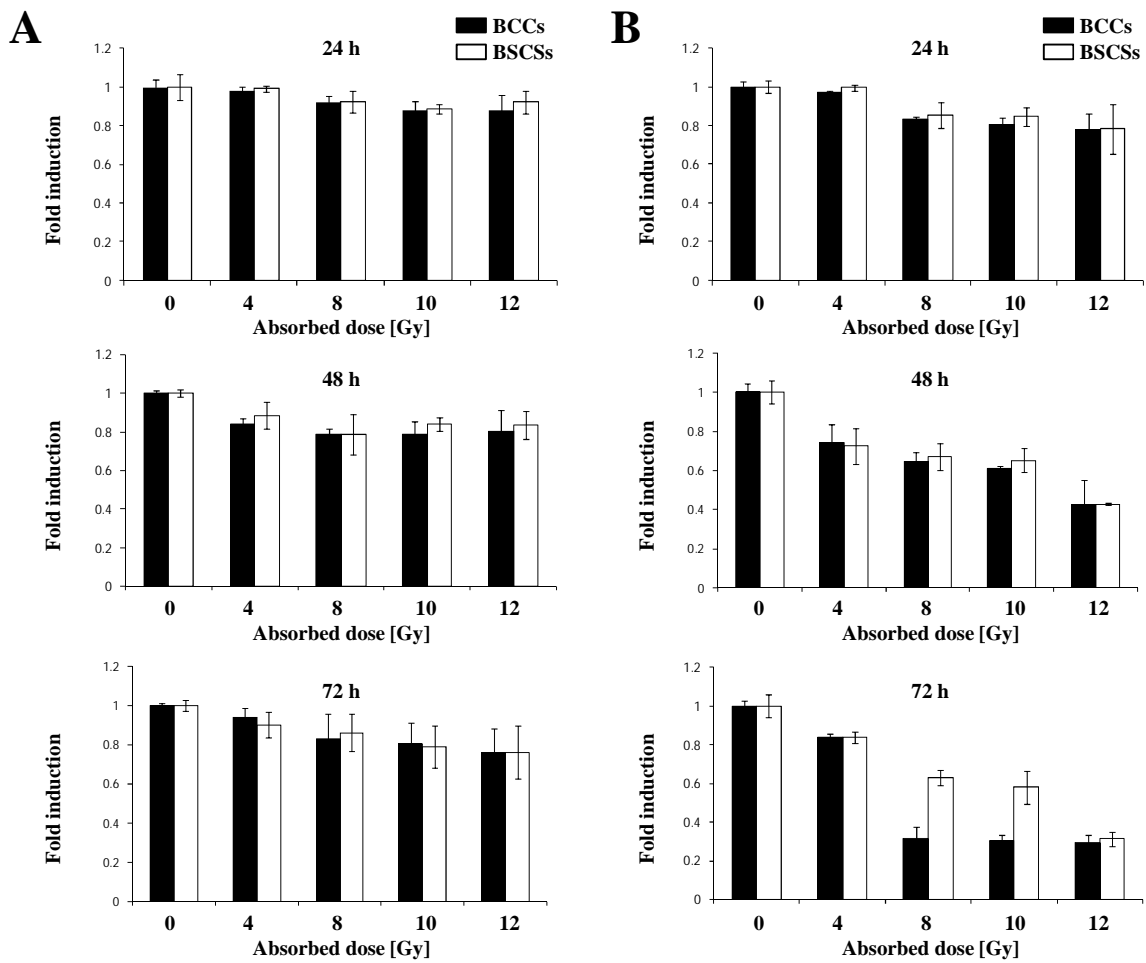

Fig. 2. Cell death following BCSC and BCC pulsed proton beam irradiation. Analysis of (A) MDA-MB-231 and (B) MCF-7 cell death 24 to 72 h after pulsed proton beam irradiation.

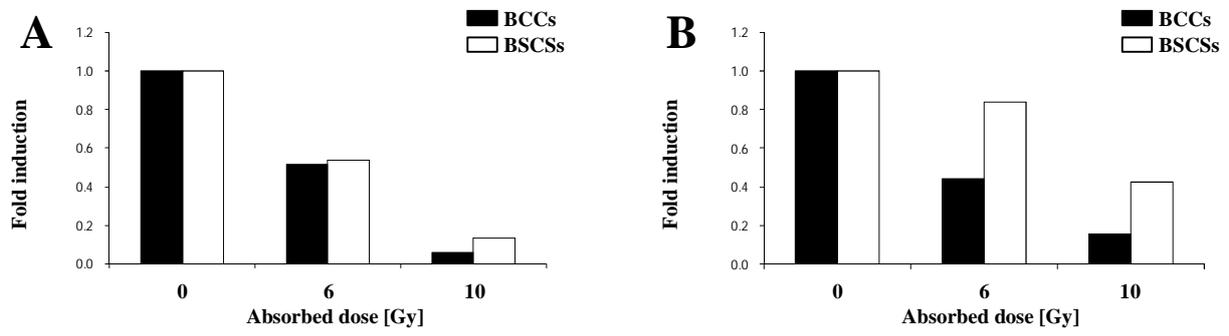

Fig. 3. BCSCs derived from MCF-7 cells are radio-resistant to pulsed proton beam irradiation. Colony formation assay results for (A) MDA-MB-231 and (B) MCF-7 cells.

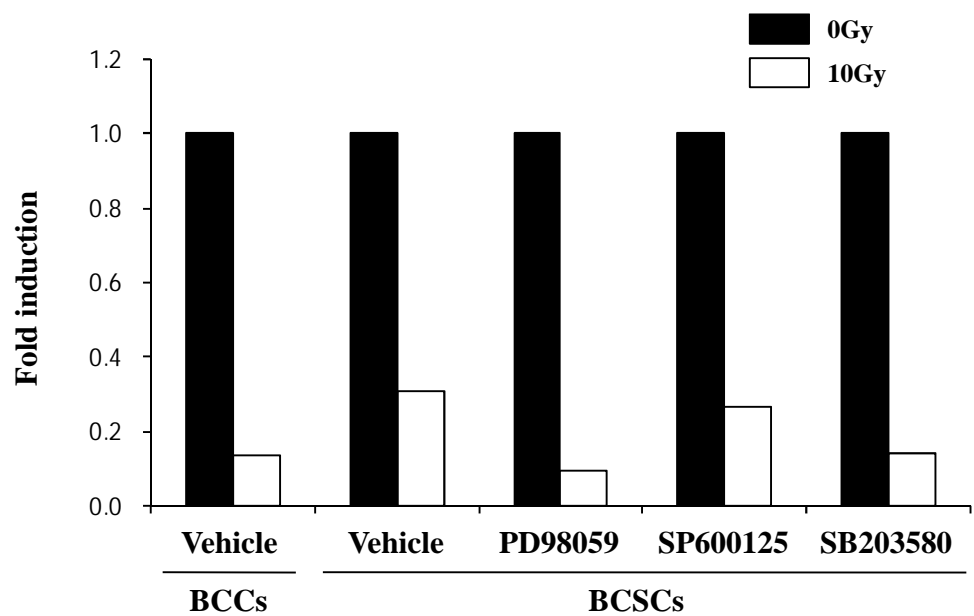

Fig. 4. Effect of ERK1/2 or p38 MAPK inhibitors on survival an after pulsed proton beam irradiation in BCSCs.